\documentclass{ws-ijmpd}

\begin{document}

\markboth{Li} {GRB physics by Fermi/LAT}

%
\catchline{}{}{}{}{}
%

\title{ON GRB PHYSICS REVEALED BY FERMI/LAT}

\author{ZHUO LI}

\address{Department of Astronomy/Kavli Institute for Astronomy and Astrophysics, Peking
University, Beijing 100871, China\\
Key Laboratory for the Structure and Evolution of Celestial Objects,
Chinese Academy of Sciences, Kunming 650011, China\\
zhuo.li@pku.edu.cn}

\maketitle


\begin{abstract}
We discuss the implications of Fermi/LAT observations on several
aspects of gamma-ray burst (GRB) physics, including the radiation
process, the emission sites, the bulk Lorentz factor, and the
pre-shock magnetic field: (1) MeV-range emission favors synchrotron
process but the highest energy ($>10$~GeV) emission may not be
synchrotron origin, more likely inverse Compton origin; (2) GRB
should have multi-zone emission region, with MeV emission produced
at smaller radii while optical and $>100$~MeV emission at larger
radii; (3) the bulk Lorentz factor can be a few $100$'s, much lower
than $10^3$, in multi-zone model; (4) the upstream magnetic field of
afterglow shock is strongly amplified to be at least in mG scale.
\end{abstract}


\section{Introduction}

After decades of research we have deeper and deeper understanding on
the nature of gamma-ray bursts (GRBs). In order to solve the
compactness problem, it is long believed that cosmological GRBs
should be produced by relativistically expanding outflow with
Lorentz factor (LF) $\Gamma>10^2$. The outflow is likely in a jet
configuration, based on the energy argument in some cases of GRBs
with huge energy $E_{iso}>10^{54}$erg, and based on observations of
light curve breaks in optical afterglows. However, beyond that very
little is known about the prompt emission, e.g., the launch of the
relativistic jet, the jet composition, the energy dissipation, and
the radiation mechanism, etc. In comparison, the afterglow phase
seems to be clearer. It is widely believed that the bulk afterglow
emission is mainly produced by a relativistic shock propagating into
the circum-burst medium, generating high energy electrons which
produce afterglow emission by synchrotron radiation. However, the
particle acceleration process in the shock is still poorly
understood.

Thanks to its high sensitivity in high energy range and wide field
of view, Fermi/LAT bringed us a brand new window to observe GRBs. I
will discuss, in my point of view, the implication of LAT
observations on GRB physics, including the radiation process, the
emission sites, the bulk LF and the pre-shock magnetic field.

\section{Main LAT Discoveries}

By the LAT observations of the past 3 years, many new features of
high energy emission in GRB prompt-burst and early-afterglow phases
were discovered (see also Ref. \refcite{overview,granot10} for an
over view). (I) In most LAT observed GRBs the onset of high energy
(LAT) emission is delayed, e.g., by few seconds for long GRBs
080916C\cite{Fermi916c} and 090902B\cite{Fermi902B} and 0.2~s for
short GRB 090510\cite{Fermi510}. (II) In general the high energy
emission is extended well beyond the end of MeV emission, e.g., up
to $10^3$s in long GRBs 080916C and 090902B and up to $10^2$s in
short GRB 090510. (III) In the brightest bursts, high energy photons
are detected up to tens of GeV, e.g., GRBs 080916C, 090510, 090902B,
and 090926A\cite{Fermi926A}. (IV) In most LAT bursts the
time-integrated spectra show single ``Band components" up to a few
10's GeV, e.g., GRB 080916C, while some cases show additional high
energy components with hard spectral indices, e.g., long GRB 090902B
and short GRB 090510. However, the $\nu F_\nu$ spectra still peak at
MeV range.

\section{Implications by LAT}

\subsection{Radiation process in prompt emission}

The huge energy released in gamma-rays by GRBs, typically
$\gtrsim10^{53}$erg, suggests that the radiation processes should be
efficient, otherwise there may be energy crisis given the limited
gravitational energy available in a stellar explosion. The most
favored radiation processes are synchrotron and inverse-Compton (IC)
radiations by electrons. There are already some expectation before
the launch of Fermi mission that there may be a high energy bump in
high energy due to IC emission. However, Fermi observations, given
its wide energy range, over 7 decades, show that most GRBs only have
single Band component, and even for those GRBs with additional high
energy components, the $\nu F_\nu$ flux peaks at MeV range. If the
GRB emission is dominated by MeV emission, the radiation process
responsible to MeV emission favors synchrotron radiation over IC
one\cite{derishev01,piran09}. This is because if the MeV photons are
produced by IC scattering of lower energy photons, the upscattering
of the MeV photons (second order IC) will produce higher energy
component with even higher flux. Ref. \refcite{wang09} analyzes GRB
080916C specifically, and concludes that if MeV emission is produced
by IC and the second order IC component is beyond the LAT energy
range (in order not conflicted with LAT observation), the required
energy should be larger than MeV emission energy by several orders
of magnitude, raising the energy crisis. Thus Fermi observations
favor {\em synchrotron process} over IC as the radiation mechanism
of MeV emission.

However, the highest energy emission may not be synchrotron origin.
The maximal synchrotron photon energy is limited by the fact that
the electron acceleration time is shorter than the synchrotron
cooling time. This argument can be transferred into a lower limit to
the bulk LF of GRB outflow. As detected by LAT, high energy emission
in bright GRBs extends to tens GeV scale with hard spectrum, it is
reasonable to believe that the spectral cutoff, if exist, should be
$\gtrsim10^2$GeV\footnote{A spectral fit to GRB 080916C spectrum
using exponential cutoff at high energy actually results in a cutoff
energy of $E_{cut}>100$~GeV, much larger than the detected highest
energy photon (B. Li and Z. Li, 2011, subm).}, which, if having
synchrotron origin, implies a bulk LF of
$\Gamma\gtrsim10^4$\cite{li10}. Such high LF may not be available
for GRB jets, then the synchrotron origin of the LAT detected
highest energy photons may not be true\cite{li10}. Other mechanisms
are required to produce the observed $\gtrsim10$GeV photons, likely
to be IC origin, e.g., IC emission from residual internal
shocks\cite{li10}.

It should be emphasized that the explanation of the GRB prompt
spectra is still an open question. In particular, the hard spectra
below break energy, especially in some GRBs with spectra harder than
$f_\nu\propto\nu^{1/3}$, are inconsistent with optically-thin
synchrotron radiation, which has been argued not in favor of
synchrotron but alternative processes, e.g., ``photosphere"
emission\cite{photosphere}. However, there are already efforts in
producing hard spectrum from synchrotron
process\cite{daigne11,nakar,derishev01}, and the detection of high
polarization in GRB prompt emission does support synchrotron
origin\cite{polarization}.

\subsection{Prompt emission sites}

LAT detects that in GRB 080916C {\em the bulk emission of the second
light-curve peak is moving toward later times as the energy
increases} (see time bin b in Fig 1 and its inset panels in Ref.
\refcite{Fermi916c}), and the time delay of $100$-MeV emission is
about 1 s relative to MeV emission. Note this is much larger than
the MeV variability timescale, $<100$~ms\cite{greiner}. In short GRB
090510, the cross correlation analysis also show that the LAT
emission is delayed relative to MeV emission, with delay time,
$\sim0.2$s, larger than the ms-scale MeV variability time by orders
of magnitude\cite{Fermi510}.

The fact that the high energy delay time is larger than low energy
(MeV) variability time by orders of magnitude, $$t_{delay}^{H}\gg
t_{var}^{L},$$ is naturally explained by multi-zone emission
regions, i.e., the low energy emission is produced at small radii,
$R_L\lesssim\Gamma^2ct_{var}^L$, while the high energy one produced
at larger radii, $R_H\sim R_L+\Gamma^2ct_{delay}^H\gg R_L$. In
addition, there is also evidence of multi-zone emission region in
optical observations. GRB 080319B shows optical time delay of $2$~s
relative to MeV emission\cite{080319b}, larger than MeV variability
time by orders of magnitude, $t_{delay}^{opt}\gg t_{var}^L$. This
also suggests that optical emission is produced at a region with
radii much larger than MeV radii, $R_{opt}\sim
R_L+\Gamma^2ct_{delay}^{opt}\gg R_L$.

The above time delay features are consistent with the residual
collisions\cite{li08,li10} in the framework of internal shock
models. After first generation collisions which produce MeV
synchrotron emission, the residual shell-shell collisions continue
taking place, smearing out the velocity fluctuation in the outflow.
The residual collisions accelerate electrons which produce longer
wavelength synchrotron emission, and generate high energy emission
by IC scattering since the outflow is under bath of the inner-origin
MeV photons\cite{li08}. The high energy delay is due to that the
higher energy photons can only avoid $\gamma\gamma$ absorption at
larger radii\cite{li10}.

The large delay time implies that one-zone (one-radius) models do
not work. However the argument does not hold in models where the
emission region is highly angularly inhomogeneous and
$R_L\lesssim\Gamma^2ct_{var}^L$ is not
necessary\cite{jet,turbulence,icmart}. But the high energy delay
still needs to be explained in these models.

\subsection{Bulk Lorentz factor}

High energy, $>10$~GeV, photons are often detected by LAT, which,
based on opacity argument, leads to more stringent constraint on
bulk LF, $\Gamma>10^3$, much larger than previously thought. If
$E_{cut}$ is much higher than the energy of observed highest energy
photons, even larger LF is required. However, this is still based on
one-zone assumption. As the high energy time delay suggests
mult-zone emission sites, this assumption should be relaxed. In a
two-zone picture, where GeV photons are produced at large radii and
attenuated by inner-originated MeV photons, the bulk LF is much
lower, i.e., a few hunderds\cite{zhao11,zou11}. In addition, based
on the residual collision model, the 1~s scale time delay of
$>100$~MeV emission in GRB 080916C suggests a typical value of bulk
LF, $\Gamma\sim300$\cite{li10}; the detection of spectral cutoff in
GRB 090926A also suggest that the LF is only a few
hundreds\cite{Fermi926A}; moreover, the geometry effect can reduce
the LF limit\cite{daigne}.

\subsection{Pre-shock magnetic field of afterglow shocks}

The LAT extended emission show a power law decay and spectral index
of about $f_\nu(t)\propto \nu^{-1}t^{-1.3}$. The light curve slope,
spectral index and flux level are consistent with the synchrotron
afterglow model. Even though it may not be true that the whole burst
is dominated by forward shock emission (as discussed in section
3.1), for the $10^3$s-scale emission the forward shock emission is
still most favored over other possible models (see a discussion in
the introduction of Ref. \refcite{wang10}).

In order for Fermi shock acceleration works, the electrons that
emitting LAT emission should survive the upstream radiative cooling
due to IC scattering the afterglow photons. This require a short
upstream residence time, i.e., the upstream magnetic field should
efficiently deflects the electron trajectory\cite{li06}. For
$>100$~MeV photons at $10^3$s scale, the upstream magnetic field is
required to satisfy $B_u>10^{0}n_0^{9/8}$mG\cite{li11}, where $n_0$
is the preshock density in unit of  $1~\rm cm^{-3}$, more stringent
than the previous constraint by X-ray afterglow observations on day
scale\cite{li06}. This suggests that the preshock magnetic field is
strongly amplified\footnote{Following the same argument
Ref.~\refcite{lowBu} results in much lower $B_u$ limit because they
assume an uniform, ordered upstream magnetic field, which optimizes
much the deflection of electrons.}, most likely by the streaming of
high energy shock accelerated particles.

\section{Summary}

LAT observations help to reveal the following facts in GRB physics:
(I) The radiation process responsible to MeV range emission
 favors synchrotron over IC radiation, however the highest
 energy emission may not be synchrotron origin, but likely IC radiation.
(II) The bulk GRB emission is multi-zone origin,
 with MeV emission produced at smallest radii while long wavelength (e.g.,
 optical) and high energy emission produced at larger radii.
(III) The bulk LF can be a few 100's if relaxing the one-zone
 assumption for the emission region.
(IV) The upstream magnetic field of afterglow shock is at least
 in mG scale, implying strong amplification of pre-shock magnetic field.

\section*{Acknowledgments}

This work is partly supported by the Foundation for the Authors of
National Excellent Doctoral Dissertations of China and the Open
Research Program of Key Laboratory for the Structure and Evolution
of Celestial Objects, Chinese Academy of Sciences.


\end{document}